\documentclass{jetpl}
\twocolumn

\usepackage{amssymb}
\usepackage{textcomp}
\usepackage{wasysym}

\lat

\title{Phase transition between ${\bf (2\times 1)}$ and ${\bf c(8\times 8)}$ reconstructions observed on the Si(001) surface around ${\bf 600^{\circ}}$C}

\rtitle{Phase transition between $(2\times 1)$ and $c(8\times 8)$ reconstructions}

\sodtitle{Phase transition between $(2\times 1)$ and $c(8\times 8)$ reconstructions observed on the Si(001) surface around $600^{\circ}$C}

\author{L.\,V.\,Arapkina\thanks[1]{e-mail: arapkina@kapella.gpi.ru},
 V.\,A.\,Yuryev\thanks[2]{http://www.gpi.ru/eng/staff\_s.php?eng=1\&id=125},
V. M. Shevlyuga, K. V.  Chizh \/}

\rauthor{Arapkina L. V., Yuryev V. A., Shevlyuga V. M., Chizh K. V.}

\sodauthor{Arapkina, Yuryev, Shevlyuga, Chizh}

\address{A.\,M.\,Prokhorov General Physics Institute of the Russian Academy of Sciences,\\ 38 Vavilov Street, Moscow, 119991, Russia}

\dates{date}{*}

\abstract{
The Si(001) surface subjected to different treatments in ultrahigh vacuum molecular beam epitaxy chamber for SiO$_2$ film decomposition    has been  {\it in situ} investigated by reflected high energy electron diffraction (RHEED) and high resolution scanning tunnelling microscopy  (STM).  A transition between $(2\times 1)$ and $(4\times 4)$ RHEED patterns was observed.  The $(4\times 4)$ pattern arose at $T \apprle 600^{\circ}$C  during sample posttreatment cooling. The reconstruction was observed to be reversible. The $c(8\times 8)$ structure was revealed by  STM  at room temperature on the same samples.  The $(4\times 4)$ patterns have been evidenced to be a manifestation of the $c(8\times 8)$ surface structure in RHEED. The phase transition appearance has been found to  depend on   thermal treatment conditions and sample cooling rate.
}


\begin{document}

\maketitle


Development of a procedure of atomically clean Si(001) surface preparation  at lowered temperatures and/or by short thermal treatments is  a keystone of creation of a
CMOS compatible process  of nanoelectronic VLSI fabrication \cite{classification,Hut_nucleation}. One of the ways of solving this problem is formation of a thin protective SiO$_2$ film   on a Si surface or surface passivation by hydrogen atoms  
during wet chemical etching with posterior silicon dioxide decomposition  or hydrogen desorption from the surface in ultrahigh vacuum (UHV) ambient \cite{cleaning_handbook,cleaning_APL}. In this connection, an issue of surface structure after these treatments becomes a task of primary importance taking into account a possible effect of  Si surface atomic-scale roughness on  formation of nanostructured elements (e.\,g., self-assembled Ge quantum dot nucleation on wetting layer in Ge/Si(001) heterostructures \cite{Hut_nucleation,Ion_irradiation_1,Smagina}).

This letter presents data of mutually complimentary investigation of  clean Si(001) surfaces prepared by different methods in an UHV molecular beam epitaxy (MBE) chamber after different processes of wet chemical etching which has been
carried out by means of high resolution scanning tunnelling microscopy (STM) and {\it in situ} reflected high energy electron diffraction (RHEED).

\begin{figure}
\begin{center}
\includegraphics[scale=1.3]{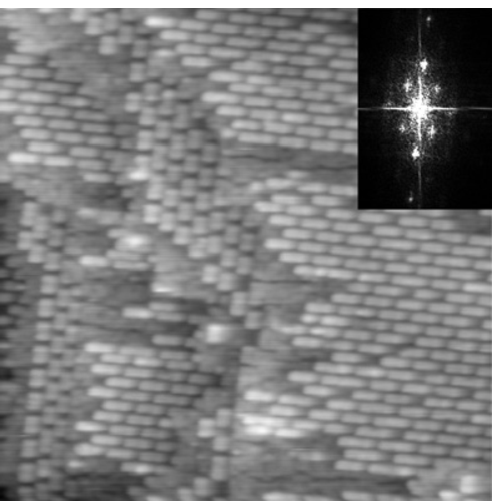}(a)
\includegraphics[scale=1.3]{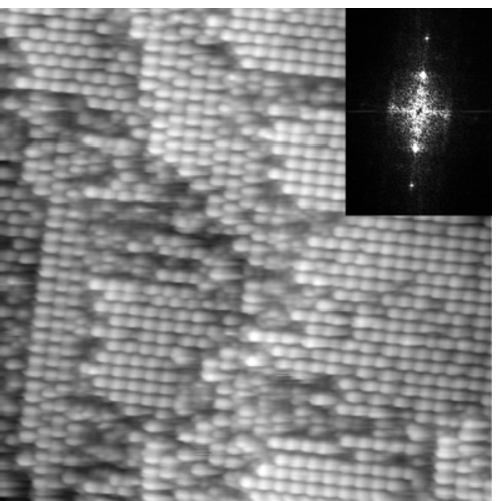}(b)
\caption{\label{fig:925C_fast}Fig.~1. 
A complimentary pair of STM images of the Si(001) surface 
after two cycles of annealing at $925^{\circ}$C for $\sim 3$ min. 
 with quenching ($50\times 60$\,nm): (a) empty state mode, $U_{\rm s}=+1.9$~V, $I_{\rm t}=80$~pA; (b) filled state mode, $U_{\rm s}=-1.5$~V, $I_{\rm t}=80$~pA; inserts present the corresponding Fourier transforms ($2.9\times 2.9$ and $3.7\times 3.7$ nm$^{-1}$, respectively). 
}
\end{center}
\end{figure}

\begin{figure}
\begin{center}
\includegraphics[scale=1.3]{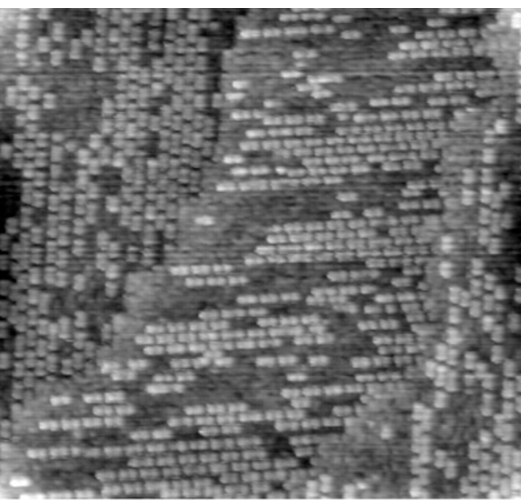}(a) 
\includegraphics[scale=1.3]{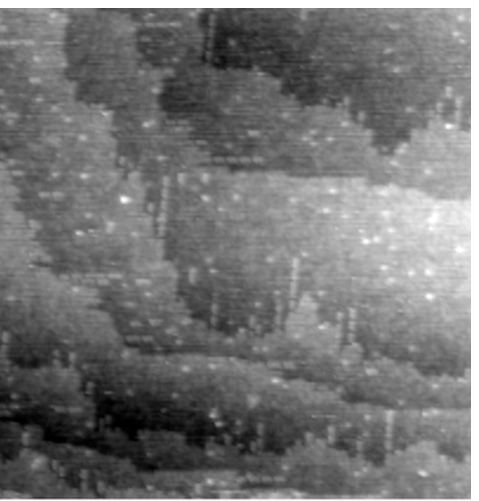}(b) 
\caption{\label{fig:925C_slow} Fig.~2. 
STM images of the Si(001) surface
after annealing at at $925^{\circ}$C for $\sim 3$ min. and slow cooling: (a) $70\times 70$\,nm, $U_{\rm s}=+1.6$~V, $I_{\rm t}=200$~pA; (b) $104\times 109$\,nm, $U_{\rm s}=+2.0$~V, $I_{\rm t}=100$~pA.
 }
\end{center}
\end{figure}

\begin{figure}
\begin{center}
\includegraphics[scale=1.3]{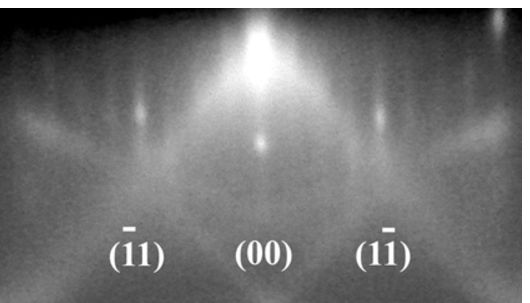}(a)
\includegraphics[scale=1.26]{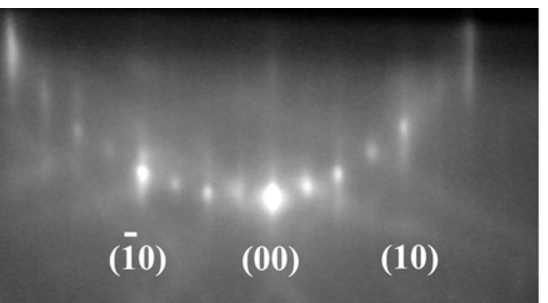}(b)
\caption{\label{fig:4x4_RHEED} Fig.~3.
Si(001)-$(4\times 4)$ RHEED patterns    obtained in (a) $[1\,1\,0]$ and (b) $[0\,1\,0]$ 
azimuths  after high-temperature annealing for SiO$_2$ film decomposition; electron beam energy is 10~keV.
}
\end{center}
\end{figure}

The experiments were made using an integrated ultra-high-vacuum (UHV) system \cite{classification} based on the Riber EVA\,32 molecular beam epitaxy chamber equipped with the Staib Instruments RH20 diffractometer of reflected high energy electrons and coupled through a transfer line with the GPI~300 UHV     scanning tunnelling microscope \cite{STM_GPI-Proc,STM_calibration}.
Samples  were 8$\times 8\times 0.4$\,mm  cut from the specially treated KDB-12 commercial B-doped    CZ Si$(100)$ wafers ($p$-type,  $\rho\,= 12~\Omega\,$cm). Initially, the specimens were chemically etched in the RCA etchant or a mixture of HNO$_3$ and HF \cite{cleaning_handbook}. The obtained samples had different thicknesses of protective SiO$_2$ films. A part of the samples, after etching in RCA, were immerged in a dilute HF solution to form a surface  passivated by hydrogen atoms (Si:H). Then, the samples (except for Si:H) were annealed at $600^{\circ}$C for not less than 6 hours at the residual gas pressure of less than $5\times 10^{-9}$~Torr; the Si:H samples were treated at the temperature of $\sim 300^{\circ}$C and the pressure of less than $5\times 10^{-11}$~Torr.

To obtain a clean Si(001) surface two standard for MBE methods of surface preparation were used:  a short high-temperature annealing and  decomposition of the SiO$_2$ film by a weak flux of Si atoms. The short  annealings were performed at 925, 800 or $650^{\circ}$C. A pressure in the MBE chamber did not exceed $5\times 10^{-9}$~Torr.
Surface cleaning was monitored by RHEED: a characteristic Si(001)-$(2\times 1)$ pattern indicated that the surface was clean. Preparation of a clean surface by silicon dioxide decomposition in Si flux was conducted at the temperature of $800^{\circ}$C and Si deposition rate of less than 0.1~\r{A}/s. The deposited Si layer thickness was greater than 30\,\r{A}. 
The samples were heated from the rear side by Ta radiators. The temperature was monitored with the IMPAC IS\,12-Si pyrometer which measured the  sample temperature through chamber windows.
Two modes of cooling was applied: quenching at the rate of $\sim 0.4$\,K/s and slow cooling at the rate of $\sim 0.17$\,K/s
The atmosphere composition in the MBE camber was monitored using the SRS~RGA-200 residual gas analyser.
The STM images were obtained in the constant tunnelling current ($I_{\rm t}$) mode at room temperature. The pressure in the STM chamber was less than  10$^{-10}$\,Torr.
The STM tip was zero-biased while the sample was positively or negatively biased ($U_{\rm s}$) when scanned in empty or filled states imaging mode. The STM tips were {\it ex situ} made of the tungsten wire and cleaned by ion bombardment \cite{W-tip}  
in a special UHV chamber connected to the STM. 
The STM images were processed afterwords using the WSxM software \cite{WSxM}.


Application of high-temperature annealings for silicon dioxide film removal can lead to appearance of a reconstruction on the Si(001) surface during wafer cooling which is different from the $(2\times 1)$ one. This reconstruction has previously been found to be $c(8\times 8)$  \cite{our_Si(001)_en}. 
This structure consists of ordered Si dimer pairs and di-vacancies which form ``rectangles'' gathered in rows running along $\langle$110$\rangle$ axes \cite{our_Si(001)_en}.
Surface coverage by this structure somewhat varies in different samples cooled at the same  rate.   Decrease of the coverage is observed with lowering of the sample cooling rate. Figs.~\ref{fig:925C_fast} and~\ref{fig:925C_slow} show STM images of the Si(001) surface obtained from samples cooled at different rates. It is seen that the same structure forms on the surfaces but with different coverages. 

Surface deoxidization has been explored during processing by means of RHEED; it has been established that according to electron diffraction patterns the $(2\times 1)$ structure forms on the surface during the high-temperature treatment at  $925^{\circ}$C. Then, on sample cooling, a phase transition goes on
which is characterized  by gradual appearance of a diffraction pattern corresponding to the $(4\times 4)$ reconstruction (Fig.~\ref{fig:4x4_RHEED}). The pattern change on screen is apparent to the eye  in the sample temperature interval from $\sim 600$ to $\sim 550^{\circ}$C. The phase transition is reversible:  recurring heating returns the $(2\times 1)$ pattern at the same temperature ($\sim 600^{\circ}$C); the $(4\times 4)$ pattern arise again on repeated cooling. Fig.~\ref{fig:925C_fast} presents STM images of a sample after two cycles of annealing and quenching. The $c(8\times 8)$ structure is seen on the images obtained in the  empty-state (Fig.~\ref{fig:925C_fast}a) and filled-state (Fig.~\ref{fig:925C_fast}a) modes; RHEED $(4\times 4)$  patterns corresponding to this surface reconstruction are presented in Fig.~\ref{fig:4x4_RHEED}. Inserts of  Fig~\ref{fig:925C_fast} show the Fourier transforms which are evidently different for different scanning modes: the empty-state image Fourier transform corresponds with  a period of 8 translations along the [110] direction; the filled-state    one exhibits a period of 4 translations in this direction. 

\begin{figure*}
\begin{center}
\includegraphics[scale=1.05]{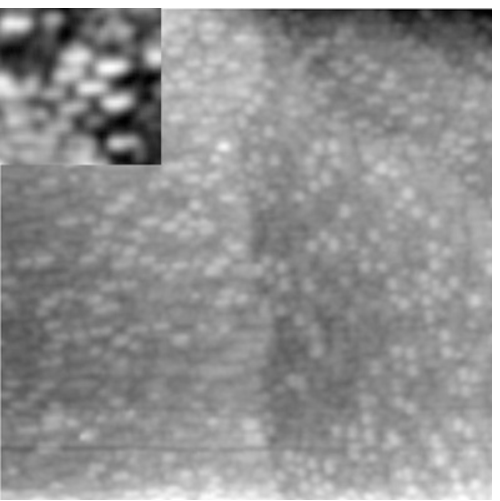}(a)
\includegraphics[scale=1.05]{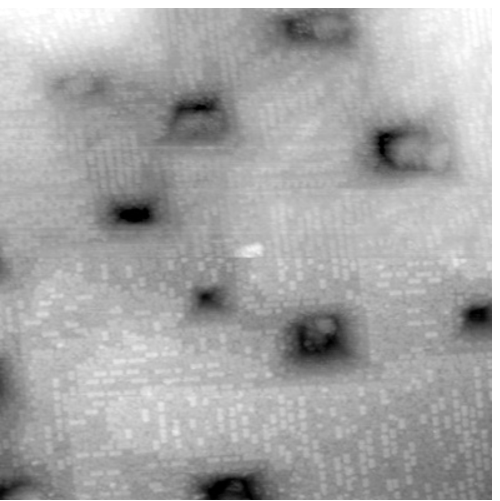}(b)
\includegraphics[scale=1.05]{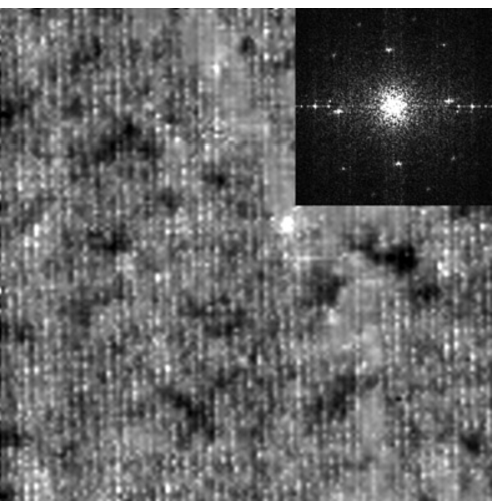}(c)
\caption{\label{fig:different_STM}Fig.~4.
STM images  of the Si(001) surface
after different treatments with posterior quenching:
(a) annealing at $800^{\circ}$C for $\sim 8$ min., 
$70\times 70$\,nm, $U_{\rm s}=+2.0$~V, $I_{\rm t}=100$~pA, an insert presents a magnified image ($11\times 11$\,nm); 
(b) annealing at $800^{\circ}$C in a weak flux of Si atoms, 
$99\times 99$\,nm, $U_{\rm s}=+1.8$~V, $I_{\rm t}=100$~pA;
(c) H atoms removal by annealing at  $650^{\circ}$C for $\sim 5$ min., 
$32.6\times 32.6$\,nm, $U_{\rm s}=+2.3$~V, $I_{\rm t}=120$~pA, an insert shows the Fourier transform ($4.6\times 4.6$ nm$^{-1}$).
 }
\end{center}
\end{figure*}

After chemical etching in RCA, samples had a thin oxide layer--- the thickness of the SiO$_2$ film in this case was $\sim 1$\,nm. The samples with thin  SiO$_2$ layers were usually annealed at $800^{\circ}$C for the oxide film removal. An STM image of such a sample annealed for 8 minutes at $800^{\circ}$C is presented in Fig.~\ref{fig:different_STM}a. The same structure as that exhibited in Figs.~\ref{fig:925C_fast} and~\ref{fig:925C_slow} is seen on the surface, but 
 much less occupancy is its characteristic feature. RHEED patterns for this surface correspond with $(2\times 1)$. 

A standard for MBE method of clean surface preparation---exposure of the surface to a weak flux of Si atoms---was utilized instead of the high-temperature annealing for oxide film decomposition from some samples with a thick SiO$_2$ film. The sample temperature during the process was $800^{\circ}$C. 
Fig.~\ref{fig:different_STM}b demonstrates an image of the Si(001) surface after such treatment. The surface reconstruction is $c(8\times 8)$ but the occupancy is not high and RHEED patterns correspond to $(2\times 1)$. Stepped rectangular pits are clearly seen in the image. Analogous pits were observed by the authors of Ref.~\cite{cleaning_APL}. They are likely one of specific defects of a surface which may be caused by etch pits arising at the phase of wet etching. 

\begin{figure}
\begin{center}
\includegraphics[scale=1.4]{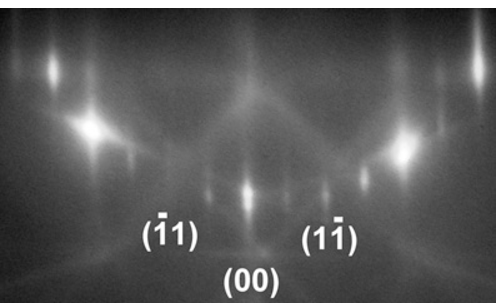}(a)
\includegraphics[scale=1.3]{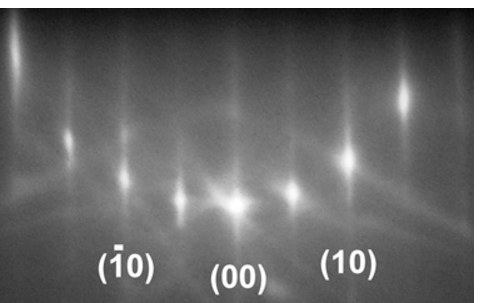}(b)
\caption{\label{fig:2x2_RHEED}Fig.~5.
Si(001)-$(2\times 2)$ RHEED patterns    obtained in (a) $[1\,1\,0]$ and (b) $[0\,1\,0]$ 
azimuths  after H atoms removal at $650^{\circ}$C; electron beam energy is 10~keV. 
}
\end{center}
\end{figure}

Samples with the surface passivated by hydrogen atoms underwent 
annealing at $650^{\circ}$C for 5 minutes.  A survey by STM revealed a $(2\times 2)$-type structure (Fig.~\ref{fig:different_STM}c): the Fourier transform of the image, which is shown in the insert, corresponds with a periodicity of 2 translations along the $\langle$110$\rangle$ directions.   Fig.~\ref{fig:2x2_RHEED} represents  RHEED patterns obtained from one of these samples with reflexes exactly matching to the reconstruction observed by STM.  The origin of this reconstruction has not been understood thus far and requires further study.

The $(2\times 1)$  RHEED pattern  was observed for the passivated samples after annealing  at  $800^{\circ}$C for 5 minutes and quenching.

Earlier, we have already presented a model of the  $c(8\times 8)$ structure formation based on the assumption that it consists of two layers \cite{our_Si(001)_en}. The uppermost layer is composed by Si ad-atoms which, at high temperatures, migrate along the underlying $(2\times 1)$ reconstructed layer. Si ad-atoms stay on the surface after the SiO$_2$ decomposition reaction: SiO$_2$\,+\,Si\,=\,2SiO\,$\uparrow$. This is additionally  confirmed by the observation of the $(2\times 1)$ reconstruction after cleaning at  $800^{\circ}$C of the surface passivated by hydrogen  because in this case Si atoms are not spent for hydrogen removal.  RHEED data may be interpreted as follows: at high temperatures, we observe a pattern produced by diffraction on the underlying layer, which corresponds to $(2 \times 1)$. As the sample is cooled atoms of the uppermost layer loose the mobility and form the $c(8\times 8)$ structure which is exhibited in diffraction patterns as  $(4\times 4)$. Possible models of atom ordering in such structure will be considered in a different article \cite{STM_RHEED}. 

A reason of the discrepancy of STM and RHEED results may be understood from the fact that the $c(8\times 8)$ structure consists of ``rectangles'' whose location is strictly predetermined by the dimer rows of the underlying layer \cite{our_Si(001)_en}. 
The dimers located on short sides of the ``rectangles'' lie on top of the lower dimer rows and are somewhat higher than the rest dimers forming the $c(8\times 8)$ structure. Fig.~\ref{fig:925C_fast}b demonstrates an STM image of the surface and its Fourier transform  which correspond to the $(4\times 4)$ reconstruction.  Depending on a magnitude of the negative bias applied to the specimen, an image can be obtained in which the ``rectangle'' looks as two parallel bars positioned  exactly on places of the extreme  dimers \cite{our_Si(001)_en}. This usually takes place at small values of the negative bias. In this case the entire visible structure corresponds to the $(4\times 4)$ reconstruction.
Thus, a cause of the discrepancy of STM and RHEED data is likely that the high energy electrons diffract on the topmost dimers, situated on sides of the ``rectangles'', rather than on different dimers of the surface structure who are somewhat lower but usually manifested in STM images except for some special mode of scanning at low negative bias when only the highest dimers contribute to the tunnelling current.


In summary, it has been established that at high-temperature processes of the silicon dioxide film removal from the Si(001) surface a reversible phase transition from  $(2\times 1)$ to  $c(8\times 8)$ structure takes place at the temperature of $\sim 600^{\circ}$C on sample cooling. The $(2\times 1)$ structure restores on heating at the same temperature. The low-temperature structure is exhibited  as $(4\times 4)$ one in  RHEED patterns.  STM images show this structure to cover different fractions of the surface area. The coverage decreases as the sample cooling rate is reduced. At small coverages,  RHEED patterns correspond to the $(2\times 1)$ reconstruction. STM images also demonstrate the $(2\times 1)$ structure in areas free of the $c(8\times 8)$ one.


The research was supported by the Education Agency of the Ministry of Education and Science of Russian Federation under the State Contract No.~$\Pi$2367.



\end{document}